\documentclass[12pt]{decar-wsd}

\usepackage{decar-common}
\usepackage{todonotes}
\usepackage{decar-post}

\AtBeginBibliography{\raggedright}

\title{Linear Matrix Inequality Approaches to Koopman Operator Approximation}
\author{Steven Dahdah\textsuperscript{1} and James Richard Forbes\textsuperscript{2}}

\begin{document}

\maketitle

\section{Introduction}

Koopman operator
theory~\cite{koopman_hamiltonian_1931, mezic_2019_spectrum,
budisic_applied_2012, mauroy_2020_koopman}
provides a means to globally represent a nonlinear system as a linear system by
transforming its states into an infinite-dimensional space of \textit{lifted
states}.
The Koopman operator advances the current lifted state of
the system to the next lifted state, much like the state transition matrix of
a linear system.
While originally proposed by B.~O.~Koopman in
1931~\cite{koopman_hamiltonian_1931}, modern computational resources, along with
recent theoretical
developments~\cite{mezic_2019_spectrum, budisic_applied_2012,
mauroy_2020_koopman},
have led to a resurgence of interest in using data-driven methods to approximate
the Koopman operator.

In general, the Koopman representation of a nonlinear system is
infinite-dimensional.
One way to approximate the Koopman operator in finite dimensions is to select a
set of lifted states and use linear regression to find a matrix approximation of
the Koopman operator, also called a \textit{Koopman
matrix}~\cite{kutz_dynamic_2016, otto_2021_koopman}.
The Koopman representation of a nonlinear system is particularly convenient for
control systems design, as its linear representation of nonlinear systems is
compatible with a wide variety of existing linear optimal control
techniques~\cite{korda_2018_linear, otto_2021_koopman, abraham_active_2019,
mamakoukas_local_2019, bruder_modeling_2019, uchida_2021_data-driven}.

The regression problem associated with finding an approximate Koopman operator
is numerically challenging, requiring regularization techniques, such as
Tikhonov regularization~\cite{tikhonov_1995_numerical} or the
lasso~\cite{tibshirani_regression_1996}, to find a suitable solution.
The novelty of this document is the reformulation of the Koopman matrix
regression problem as a convex optimization problem with linear matrix
inequality (LMI) constraints and the use of additional LMIs to, for
instance, regularize the optimization problem.
In particular, regularizers with LMI forms, such as the matrix two-norm or the
\Hinf{} norm, can be added to the optimization problem in a modular fashion.
Additional stability constraints can also be added in the same way. Although
convex optimization and LMIs have previously been used to synthesize controllers
for Koopman models~\cite{uchida_2021_data-driven}, these tools have not yet been
leveraged when solving the regression problem associated with finding the
Koopman matrix.
In~\cite{sznaier_2021_convex}, a related optimization problem is posed where
both the Koopman matrix and lifting functions are unknown. While the problem is
NP-hard, a convex relaxation allows both to be found by solving two semidefinite
programs.

A particular novelty of this document is solving the Koopman regression problem
with a system norm regularizer.
Although this document explores the use of the
\Hinf{}~norm~\cite[\S3.2]{caverly_2019_lmi} as a regularizer, any system norm
can be used, such as the
\Htwo{}~norm~\cite[\S3.3]{caverly_2019_lmi} or a mixed
\Htwo{}~norm~\cite[\S3.5]{caverly_2019_lmi}.
This systems perspective on the regression problem is a natural fit with the
Koopman operator because the Koopman matrix describes the time evolution of the
data associated with a dynamic system.
While the \Hinf{}~norm of the Koopman operator has previously been considered
in~\cite{hara_2020_learning}, it is in the form of a hard constraint on the
system's dissipativity.
The use of a system norm to regularize an optimization problem enables a
systems interpretation of the entire regularization procedure.
For example, using a system norm as a regularizer enables the use
of weighting functions that can explicitly penalize system gain in a particular
frequency band.

This document focuses on the formulation of the Koopman matrix regression
problem using convex optimization and LMIs, and demonstrates how LMIs can be
leveraged to regularize or enforce additional constraints. This document does
not present any numerical results, which are ongoing and will be continued in
the future.

\section{Koopman operator background}

\subsection{Koopman operator theory}

Consider the discrete-time nonlinear process
\begin{equation}
    \mbf{x}_{k+1} = \mbf{f}(\mbf{x}_{k}), \label{eq:dyn_noin}
\end{equation}
where $\mbf{x}_{k} \in \mc{M}$ evolves on a smooth
manifold $\mc{M} \subseteq \mathbb{R}^{m \times 1}$, which is often just the entirety of
$\mathbb{R}^{m \times 1}$.

Let $\psi: \mc{M} \to \mathbb{R}$ be a \textit{lifting function}, where $\psi
\in \mc{H}$. Any function of $\mbf{x}_{k}$ that returns a scalar is a lifting
function. There are therefore infinitely many lifting functions, and they form
a Hilbert space $\mc{H}$. The \textit{Koopman operator} $\mc{U}: \mc{H} \to
\mc{H}$ is a linear operator that advances all scalar-valued lifting functions
in time by one timestep. That is~\cite[\S3.2]{kutz_dynamic_2016},
\begin{equation}
    (\mc{U} \psi)(\cdot) = (\psi \circ \mbf{f})(\cdot). \label{eq:koopman_def_noin}
\end{equation}
Using the Koopman operator, the dynamics of~\eqref{eq:dyn_noin} may then be
rewritten linearly in terms of $\psi$ as
\begin{equation}
    \psi(\mbf{x}_{k+1}) = (\mc{U} \psi)(\mbf{x}_{k}). \label{eq:koopman_dyn}
\end{equation}
In finite dimensions,~\eqref{eq:koopman_dyn} is approximated by
\begin{equation}
    \mbs{\psi}(\mbf{x}_{k+1}) = \mbf{U} \mbs{\psi}(\mbf{x}_{k}) + \mbf{r}_k,%
    \label{eq:koopman_approx_noin}
\end{equation}
where $\mbs{\psi}: \mc{M} \to \mathbb{R}^{p \times 1}$, $\mbf{U} \in
\mathbb{R}^{p \times p}$, and $\mbf{r}_k$ is the residual error. Since each
element of $\mbs{\psi}$ is a member of $\mc{H}$, $\mbs{\psi}$ is called a
\textit{vector-valued lifting function}. The \textit{Koopman matrix} $\mbf{U}$
is a matrix approximation of the Koopman operator.

\subsection{Koopman operator theory with inputs}

If the discrete-time nonlinear process has exogenous inputs, the definitions of
the lifting functions and Koopman operator must be adjusted. Consider
\begin{equation}
    \mbf{x}_{k+1} = \mbf{f}(\mbf{x}_{k}, \mbf{u}_k), \label{eq:dyn_in}
\end{equation}
where $\mbf{x}_{k} \in \mc{M} \subseteq \mathbb{R}^{m \times 1}$ and
$\mbf{u}_{k} \in \mc{N} \subseteq \mathbb{R}^{n \times 1}$.

In this case, the lifting functions become
$\psi: \mc{M} \times \mc{N} \to \mathbb{R},\ \psi \in \mc{H}$
and the Koopman operator $\mc{U}: \mc{H} \to \mc{H}$ is instead defined so that
\begin{equation}
    (\mc{U} \psi)(\mbf{x}_{k}, \mbf{u}_{k})
    = \psi(\mbf{f}(\mbf{x}_{k}, \mbf{u}_k), \star),
    \label{eq:koop-def-in}
\end{equation}
where $\star=\mbf{u}_k$ if the input has state-dependent dynamics, or
$\star=\mbf{0}$ if the input has no dynamics~\cite[\S6.5]{kutz_dynamic_2016}.
The input is state-dependent if it is computed by a controller.

Let the vector-valued lifting function
$\mbs{\psi}: \mc{M} \times \mc{N} \to \mathbb{R}^{p \times 1}$ be partitioned
as
\begin{equation}
    \mbs{\psi}(\mbf{x}_k, \mbf{u}_k) = \begin{bmatrix}
        \mbs{\vartheta}(\mbf{x}_k) \\
        \mbs{\upsilon}(\mbf{x}_k, \mbf{u}_k)
    \end{bmatrix},
\end{equation}
where $\mbs{\vartheta}: \mc{M} \to \mathbb{R}^{p_\vartheta \times 1}$,
$\mbs{\upsilon}: \mc{M} \times \mc{N} \to \mathbb{R}^{p_\upsilon \times 1}$,
and $p_\vartheta + p_\upsilon = p$.
In the case where the input has no dynamics,~\eqref{eq:koop-def-in} has
the form~\cite[\S6.5.1]{kutz_dynamic_2016}
\begin{equation}
    \mbs{\vartheta}(\mbf{x}_{k+1}) \\
    =
    \mbf{U}
    \mbs{\psi}(\mbf{x}_{k}, \mbf{u}_{k})
    + \mbf{r}_k,
\end{equation}
where
\begin{equation}
    \mbf{U} =
    \begin{bmatrix}
        \mbf{A} & \mbf{B}
    \end{bmatrix}.
\end{equation}
When expanded, this yields the familiar linear state space form,
\begin{equation}
    \mbs{\vartheta}(\mbf{x}_{k+1})
    =
    \mbf{A} \mbs{\vartheta}(\mbf{x}_{k})
    + \mbf{B} \mbs{\upsilon}(\mbf{x}_k, \mbf{u}_k)
    + \mbf{r}_k.
    \label{eq:dyn_approx_ss_in}
\end{equation}

\subsection{Approximating the Koopman operator}

To approximate the Koopman matrix from data, consider a
dataset $\mc{D} = {\{\mbf{x}_k, \mbf{u}_k\}}_{k=0}^q$ and the corresponding
lifted snapshot matrices
\begin{align}
    \mbs{\Psi} &= \begin{bmatrix}
        \mbs{\psi}_{0} & \mbs{\psi}_{1} & \cdots & \mbs{\psi}_{q-1}
    \end{bmatrix} \in \mathbb{R}^{p \times q}, \\
    \mbs{\Theta}_+ &= \begin{bmatrix}
        \mbs{\vartheta}_{1} & \mbs{\vartheta}_{2} & \cdots & \mbs{\vartheta}_{q}
    \end{bmatrix} \in \mathbb{R}^{p_\vartheta \times q}, \label{eq:Theta}
\end{align}
where $\mbs{\psi}_k = \mbs{\psi}(\mbf{x}_k, \mbf{u}_k)$ and $\mbs{\vartheta}_k
= \mbs{\vartheta}(\mbf{x}_k)$. Note that time-shifted input snapshots are not
required.

The Koopman matrix that minimizes
\begin{equation}
    J(\mbf{U}) = \|\mbs{\Theta}_+ - \mbf{U} \mbs{\Psi}\|_\frob^2
    \label{eq:koopman-cost}
\end{equation}
is therefore~\cite[\S1.2.1]{kutz_dynamic_2016}
\begin{equation}
    \mbf{U} = \mbs{\Theta}_+ \mbs{\Psi}^\dagger,
    \label{eq:koopman-soln}
\end{equation}
where ${(\cdot)}^\dagger$ denotes the Moore-Penrose pseudoinverse.

\subsection{Extended DMD}

Extended Dynamic Mode Decomposition (EDMD)~\cite{williams_data-driven_2015} is a
method to compute~\eqref{eq:koopman-soln} that reduces the computational cost
when the number of snapshots is much larger than the dimension of the lifted
state (\ie{}, $p \ll q$)~\cite[\S10.3.1]{kutz_dynamic_2016}. Specifically, it
reduces the size of the pseudoinverse required.

Consider the least-squares solution for the Koopman matrix,
\begin{align}
    \mbf{U}
    &=
    \mbs{\Theta}_+ \mbs{\Psi}^\dagger \\
    &=
    \mbs{\Theta}_+ \mbf{1} \mbs{\Psi}^\dagger \\
    &=
    \mbs{\Theta}_+
    (\mbs{\Psi}^\trans \mbs{\Psi}^{\trans^\dagger})
    \mbs{\Psi}^\dagger \\
    &=
    (\mbs{\Theta}_+ \mbs{\Psi}^\trans)
    {(\mbs{\Psi} \mbs{\Psi}^\trans)}^\dagger \\
    &=
    \mbf{G}
    \mbf{H}^\dagger,
    \label{eq:edmdi-sol}
\end{align}
where
\begin{align}
    \mbf{G}
    &=
    \frac{1}{q}
    \mbs{\Theta}_+ \mbs{\Psi}^\trans \in \mathbb{R}^{p_\vartheta \times p},
    \label{eq:edmdi-sol-1-scaled}
    \\
    \mbf{H}
    &=
    \frac{1}{q}
    \mbs{\Psi} \mbs{\Psi}^\trans \in \mathbb{R}^{p \times p}.
    \label{eq:edmdi-sol-2-scaled}
\end{align}

Since $p \ll q$, EDMD greatly reduces the dimension of the pseudo-inverse
operation required to compute $\mbf{U}$~\cite[\S10.3.1]{kutz_dynamic_2016}.
To improve numerical conditioning, $\mbf{G}$ and $\mbf{H}$ are often scaled by
the number of snapshots $q$, as in~\eqref{eq:edmdi-sol-1-scaled}
and~\eqref{eq:edmdi-sol-2-scaled}.

\subsection{Extended DMD with Tikhonov regularization}

Tikhonov regularization~\cite{tikhonov_1995_numerical, james_introduction_2013},
which penalizes the Frobenius norm of the unknown matrix in a linear regression
problem, can be used to improve the condition number of $\mbf{H}$
in~\eqref{eq:edmdi-sol-2-scaled}.  Consider the regularized EDMD cost function,
\begin{align}
    J(\mbf{U}; \alpha)
    &=
    \frac{1}{q}
    \|\mbs{\Theta}_+ - \mbf{U}\mbs{\Psi}\|_\frob^2
    + \frac{\alpha}{q}
    \|\mbf{U}\|_\frob^2
    \\
    &=
    \frac{1}{q}
    \trace{\left(%
        (\mbs{\Theta}_+ - \mbf{U}\mbs{\Psi})
        {(\mbs{\Theta}_+ - \mbf{U}\mbs{\Psi})}^\trans
    \right)}
    +
    \frac{\alpha}{q}
    \trace{\left(\mbf{U}\mbf{U}^\trans\right)}
    \\
    &=
    \trace{\left(%
        \frac{1}{q}
        \mbs{\Theta}_+ \mbs{\Theta}_+^\trans
        -
        \mbf{G}
        \mbf{U}^\trans
        -
        \mbf{U}
        \mbf{G}^\trans
        +
        \mbf{U}
        \left(%
            \mbf{H}
            +
            \frac{\alpha}{q}
            \mbf{1}
        \right)
        \mbf{U}^\trans
    \right)}.%
    \label{eq:tik-regularized-edmd-1}
\end{align}
Comparing~\eqref{eq:tik-regularized-edmd-1} with
the unregularized cost function, given by
\begin{equation}
    J(\mbf{U})
    =
    \trace{\left(%
        \frac{1}{q}
        \mbs{\Theta}_+ \mbs{\Theta}_+^\trans
        -
        \mbf{G}
        \mbf{U}^\trans
        -
        \mbf{U}
        \mbf{G}^\trans
        +
        \mbf{U}
        \mbf{H}
        \mbf{U}^\trans
    \right)},\label{eq:tik-regularized-edmd-2}
\end{equation}
demonstrates that
\begin{equation}
    \mbf{U}
    =
    \mbf{G}
    \mbf{H}_\alpha^\dagger
\end{equation}
minimizes the Tikhonov-regularized cost function, where
\begin{equation}
    \mbf{H}_\alpha
    =
    \mbf{H} + \frac{\alpha}{q}\mbf{1}.%
    \label{eq:H_alpha}
\end{equation}

\section{Reformulating EDMD using LMIs}%
\label{sec:lmi}

\subsection{Reformulating the regression problem}

To add other regularizers in a modular fashion, the Koopman operator regression
problem is reformulated as a convex optimization problem with LMI constraints.
Recall that the Koopman matrix $\mbf{U}$ minimizes
\begin{equation}
    J(\mbf{U}) = \|\mbs{\Theta}_+ - \mbf{U} \mbs{\Psi}\|_\frob^2.
\end{equation}
This cost function can be rewritten as a convex optimization problem with
linear matrix inequality (LMI) constraints. Specifically, consider
\begin{align}
    J(\mbf{U})
    &=
    \frac{1}{q}
    \|\mbs{\Theta}_+ - \mbf{U} \mbs{\Psi}\|_\frob^2
    \\
    &=
    \frac{1}{q}
    \trace{\left(
        \left(\mbs{\Theta}_+ - \mbf{U} \mbs{\Psi}\right)
        {\left(\mbs{\Theta}_+ - \mbf{U} \mbs{\Psi}\right)}^\trans
    \right)}
    \\
    &=
    \frac{1}{q}
    \trace{\left(
        \mbs{\Theta}_+ \mbs{\Theta}_+^\trans
    \right)}
    -
    \frac{1}{q}
    \trace{\left(
        \mbs{\Theta}_+
        \mbs{\Psi}^\trans
        \mbf{U}^\trans
        +
        \mbf{U}
        \mbs{\Psi}
        \mbs{\Theta}_+^\trans
    \right)}
    +
    \frac{1}{q}
    \trace{\left(
        \mbf{U}
        \mbs{\Psi} \mbs{\Psi}^\trans
        \mbf{U}^\trans
    \right)}
    \\
    &=
    \underbrace{%
        \frac{1}{q}
        \trace{\left(
            \mbs{\Theta}_+ \mbs{\Theta}_+^\trans
        \right)}
    }_{c}
    -
    2 \trace{\bigg(
        \mbf{U}
        \underbrace{%
            \left(
                \frac{1}{q}
                \mbs{\Psi}
                \mbs{\Theta}_+^\trans
            \right)
        }_{\mbf{G}^\trans}
    \bigg)}
    +
    \trace{\bigg(
        \mbf{U}
        \underbrace{%
            \left(
                \frac{1}{q}
                \mbs{\Psi} \mbs{\Psi}^\trans
            \right)
        }_{\mbf{H}}
        \mbf{U}^\trans
    \bigg)}
    \\
    &=
    c
    -
    2 \trace{\left(
        \mbf{U} \mbf{G}^\trans
    \right)}
    +
    \trace{\left(
        \mbf{U} \mbf{H} \mbf{U}^\trans
    \right)}, \label{eq:lmi-int-cost}
\end{align}
where $c$ is a scalar constant,
$\mbf{G}$ is defined in~\eqref{eq:edmdi-sol-1-scaled},
and
$\mbf{H}$ is defined in~\eqref{eq:edmdi-sol-2-scaled}.

The minimization of~\eqref{eq:lmi-int-cost} is equivalent to the minimization of
\begin{equation}
    J(\mbf{U}, \mbf{\nu}, \mbf{W})
    =
    c
    -
    2 \trace{\left(
        \mbf{U} \mbf{G}^\trans
    \right)}
    +
    \nu
\end{equation}
subject to
\begin{align}
    \trace{(\mbf{W})} &< \nu,
    \\
    \mbf{W} &> 0,
    \\
    \mbf{U}\mbf{H}\mbf{U}^\trans &< \mbf{W},\label{eq:pre-schur}
\end{align}
where $\nu$ and $\mbf{W}$ are slack variables that allow the cost function to be
rewritten using LMIs~\cite[\S2.15.1]{caverly_2019_lmi}. Using the Schur
complement~\cite[\S2.3.1]{caverly_2019_lmi}, the quadratic
term~\eqref{eq:pre-schur} can be rewritten as
\begin{align}
    \mbf{W} - \mbf{U}\mbf{H}\mbf{U}^\trans > 0
    \iff
    \begin{bmatrix}
        \mbf{W} & \mbf{U} \\
        \mbf{U}^\trans & \mbf{H}^{-1}
    \end{bmatrix}
    > 0,\ \mbf{H} > 0.
\end{align}
Note that $\mbf{H} = \mbf{H}^\trans > 0$ if the columns of $\mbs{\Psi}$ are
linearly independent.
Minimizing~\eqref{eq:koopman-cost} is therefore equivalent to
\begin{align}
    \min\;&
    J(\mbf{U}, \mbf{\nu}, \mbf{W})
    =
    c
    -
    2 \trace{\left(
        \mbf{U} \mbf{G}^\trans
    \right)}
    +
    \nu
    \\
    \mathrm{s.t.}\;&
    \trace{(\mbf{W})} < \nu,
    \\
    &\mbf{W} > 0,
    \\
    &\begin{bmatrix}
        \mbf{W} & \mbf{U} \\
        \mbf{U}^\trans & \mbf{H}^{-1}
    \end{bmatrix}
    > 0.
    \label{eq:inverted-h}
\end{align}
Both the objective function and constraints are convex, and $\mbf{U}$ appears
linearly in all of them.

\subsection{Introducing Tikhonov regularization}

Recall the Tikhonov-regularized cost function,
\begin{equation}
    J(\mbf{U}; \alpha)
    =
    \|\mbs{\Theta}_+ - \mbf{U} \mbs{\Psi}\|_\frob^2
    +
    \alpha
    \|\mbf{U}\|_\frob^2.
\end{equation}
Its equivalent LMI form is
\begin{align}
    \min\;&
    J(\mbf{U}, \mbf{\nu}, \mbf{W}; \alpha)
    =
    c
    -
    2 \trace{\left(
        \mbf{U} \mbf{G}^\trans
    \right)}
    +
    \nu
    \\
    \mathrm{s.t.}\;&
    \trace{(\mbf{W})} < \nu,
    \\
    &\mbf{W} > 0,
    \\
    &\begin{bmatrix}
        \mbf{W} & \mbf{U} \\
        \mbf{U}^\trans & \mbf{H}_\alpha^{-1}
    \end{bmatrix}
    > 0,
    \label{eq:inverted-h-tikhonov}
\end{align}
where, $\mbf{H}$ has simply been replaced with $\mbf{H}_\alpha$
in~\eqref{eq:inverted-h-tikhonov}.

\subsection{Avoiding matrix inversion}

Computing the inverse of $\mbf{H}_\alpha$ in~\eqref{eq:inverted-h-tikhonov} is
numerically problematic and can be avoided using a matrix decomposition to split
$\mbf{H}_\alpha$ into
\begin{equation}
    \mbf{H}_\alpha
    =
    \mbf{L}_\alpha^{}
    \mbf{L}_\alpha^\trans.
\end{equation}
The matrix $\mbf{L}_\alpha$ can be found using a Cholesky decomposition or
eigendecomposition of $\mbf{H}_\alpha$, or a singular value decomposition of
$\mbs{\Psi}$.
Assuming this decomposition has been performed, the quadratic term in the
optimization problem becomes
\begin{align}
    \mbf{W} - \mbf{U}\mbf{H}_\alpha\mbf{U}^\trans
    &=
    \mbf{W} - \mbf{U} \mbf{L}_\alpha^{} \mbf{L}_\alpha^\trans \mbf{U}^\trans
    \\
    &=
    \mbf{W} - \left(%
        \mbf{U} \mbf{L}_\alpha
    \right)
    \mbf{1}
    {\left(%
        \mbf{U} \mbf{L}_\alpha
    \right)}^\trans.
\end{align}
Applying the Schur complement~\cite[\S2.3.1]{caverly_2019_lmi} once again yields
a new form of~\eqref{eq:inverted-h-tikhonov},
\begin{equation}
    \begin{bmatrix}
        \mbf{W} & \mbf{U} \mbf{L}_\alpha^{} \\
        \mbf{L}_\alpha^\trans \mbf{U}^\trans & \mbf{1}
    \end{bmatrix}
    >
    0.
\end{equation}
This form trades off a matrix inverse for a matrix decomposition. The new
optimization problem without matrix inversion is
\begin{align}
    \min\;&
    J(\mbf{U}, \mbf{\nu}, \mbf{W}; \alpha)
    =
    c
    -2 \trace{\left(
        \mbf{U} \mbf{G}^\trans
    \right)}
    +
    \nu
    \\
    \mathrm{s.t.}\;&
    \trace{(\mbf{W})} < \nu,
    \\
    &\mbf{W} > 0,
    \\
    &\begin{bmatrix}
        \mbf{W} & \mbf{U} \mbf{L}_\alpha^{} \\
        \mbf{L}_\alpha^\trans \mbf{U}^\trans & \mbf{1}
    \end{bmatrix}
    >
    0,
    \label{eq:noninverted-h}
\end{align}
where $\mbf{H}_\alpha = \mbf{L}_\alpha^{}\mbf{L}_\alpha^\trans$.
This formulation of the optimization problem is almost always preferable to the
formulation that requires inverting $\mbf{H}_\alpha$.

Two possible methods to compute $\mbf{L}_\alpha$ are presented, but any suitable
matrix decomposition can be used. Using the eigendecomposition
\begin{align}
    \mbf{H}_\alpha
    &=
    \mbf{V}
    \mbs{\Lambda}
    \mbf{V}^\trans
    \\
    &=
    \mbf{V}
    \sqrt{\mbs{\Lambda}}
    \sqrt{\mbs{\Lambda}}
    \mbf{V}^\trans,
\end{align}
it follows that
\begin{equation}
    \mbf{L}_\alpha
    =
    \mbf{V}\sqrt{\mbs{\Lambda}}.
\end{equation}
Another option is to leverage the singular value decomposition of $\mbs{\Psi}$,
\begin{equation}
    \mbs{\Psi}
    =
    \mbf{Q}
    \mbs{\Sigma}
    \mbf{Z}^\trans.
\end{equation}
Recalling that $\mbf{H} = \frac{1}{q}\mbs{\Psi}\mbs{\Psi}^\trans$, the matrix
$\mbf{H}_\alpha$ then becomes
\begin{align}
    \mbf{H}_\alpha
    &=
    \frac{1}{q}
    \mbf{Q}\mbs{\Sigma}^2\mbf{Q}^\trans
    +
    \frac{\alpha}{q}\mbf{1}
    \\
    &=
    \frac{1}{q}
    \mbf{Q}
    \mbs{\Sigma}^2
    \mbf{Q}^\trans
    +
    \mbf{Q}
    \left(%
        \frac{\alpha}{q}\mbf{1}
    \right)
    \mbf{Q}^\trans
    \\
    &=
    \mbf{Q}
    \left(%
        \frac{1}{q}
        \mbs{\Sigma}^2
        +
        \frac{\alpha}{q}\mbf{1}
    \right)
    \mbf{Q}^\trans.
\end{align}
The matrix $\mbf{L}_\alpha$ is therefore
\begin{equation}
    \mbf{L}_\alpha
    =
    \mbf{Q}
    \sqrt{%
        \frac{1}{q}
        \mbs{\Sigma}^2
        +
        \frac{\alpha}{q}\mbf{1}
    }.\label{eq:svd-L}
\end{equation}
Note that the matrix square root in~\eqref{eq:svd-L} is easy to compute, as its
radicand is diagonal.

Tikhonov regularization can be added to any cost function in this document by
substituting
$\mbf{H}$ and $\mbf{L}$ for $\mbf{H}_\alpha$ for $\mbf{L}_\alpha$.
This allows for mixed regularization in the style of the elastic
net~\cite{zou_2005_regularization}.
However, for the sake of brevity, this is not shown.

\section{Matrix norm regularization}

\subsection{Matrix two-norm regularization}

Matrix two-norm regularization has an LMI form that can be easily incorporated
into the optimization problem.
The regularized cost function is
\begin{equation}
    J(\mbf{U}; \beta)
    =
    \|\mbs{\Theta}_+ - \mbf{U} \mbs{\Psi}\|_\frob^2
    +
    \beta \|\mbf{U}\|_2,
    \label{eq:twonorm-reg-cost}
\end{equation}
where $\beta$ is the regularization coefficient. The matrix two-norm of a
matrix is its maximum singular value. That is,
\begin{align}
    \|\mbf{U}\|_2
    &=
    \sqrt{\bar{\lambda}\left(\mbf{U}^\trans\mbf{U}\right)}
    \\
    &=
    \bar{\sigma}(\mbf{U}),
\end{align}
where
$\bar{\lambda}(\cdot)$ is the maximum eigenvalue and
$\bar{\sigma}(\cdot)$ is the maximum singular value.

Consider the modified optimization problem
\begin{align}
    \min\;&
    J(\mbf{U}, \gamma; \beta)
    =
    \frac{1}{q}
    \|\mbs{\Theta}_+ - \mbf{U} \mbs{\Psi}\|_\frob^2
    +
    \frac{\beta}{q}
    \gamma
    \\
    \mathrm{s.t.}\;&
    {\bar{\sigma}(\mbf{U})} < \gamma.
    \label{eq:max-sing-val-constraint}
\end{align}
The constraint~\eqref{eq:max-sing-val-constraint} can be rewritten
as~\cite[\S2.11.1]{caverly_2019_lmi}
\begin{equation}
    \begin{bmatrix}
        \gamma \mbf{1} & \mbf{U} \\
        \mbf{U}^\trans & \gamma \mbf{1}
    \end{bmatrix}
    >
    0.
\end{equation}
It follows that the optimization problem
\begin{align}
    \min\;&
    J(\mbf{U}, \mbf{\nu}, \mbf{W}, \gamma; \beta)
    =
    c
    -2 \trace{\left(
        \mbf{U} \mbf{G}^\trans
    \right)}
    +
    \nu
    +
    \frac{\beta}{q} \gamma
    \\
    \mathrm{s.t.}\;&
    \trace{(\mbf{W})} < \nu,
    \\
    &\mbf{W} > 0,
    \\
    &\begin{bmatrix}
        \mbf{W} & \mbf{U} \mbf{L} \\
        \mbf{L}^\trans \mbf{U}^\trans & \mbf{1}
    \end{bmatrix}
    >
    0,
    \\
    &\begin{bmatrix}
        \gamma \mbf{1} & \mbf{U} \\
        \mbf{U}^\trans & \gamma \mbf{1}
    \end{bmatrix}
    >
    0,
\end{align}
where $\mbf{H} = \mbf{L} \mbf{L}^\trans$,
is equivalent to minimizing~\eqref{eq:twonorm-reg-cost}.

\subsection{Nuclear norm regularization}

Nuclear norm regularization~\cite{recht_guaranteed_2010, blomberg_2016_nuclear}
can be incorporated to favour low-rank Koopman operators. The regularized cost
function is
\begin{equation}
    J(\mbf{U}; \beta)
    =
    \|\mbs{\Theta}_+ - \mbf{U} \mbs{\Psi}\|_\frob^2
    +
    \beta \|\mbf{U}\|_*,
    \label{eq:nuclear-reg-cost}
\end{equation}
where $\beta$ is the regularization coefficient.
The nuclear norm of a matrix is defined as
\begin{align}
    \|\mbf{U}\|_*
    &=
    \trace{\left(
        \sqrt{\mbf{U}^\trans \mbf{U}}
    \right)}
    \\
    &=
    \sum_{i=0}^{p_\vartheta - 1} \sigma_i(\mbf{U}),
\end{align}
where $\sigma_i(\cdot)$ is the $i$th singular value. Recall that
$\mbf{U} \in \mathbb{R}^{p_\vartheta \times p}$.
The solution to the optimization problem
\begin{equation}
    \min\;
    \|\mbf{U}\|_*
\end{equation}
is equivalent to the solution to the optimization
problem~\cite{recht_guaranteed_2010}~\cite[\S2.11.6]{caverly_2019_lmi}
\begin{align}
    \min\;&
    \frac{1}{2}\left(
        \trace{(\mbf{V}_1)}
        +
        \trace{(\mbf{V}_2)}
    \right)
    \\
    \mathrm{s.t.}\;&
    \begin{bmatrix}
        \mbf{V}_1 & \mbf{U} \\
        \mbf{U}^\trans & \mbf{V}_2
    \end{bmatrix}
    \geq 0,
\end{align}
where
$\mbf{V}_1 = \mbf{V}_1^\trans$
and
$\mbf{V}_2 = \mbf{V}_2^\trans$.
It follows that the optimization problem
\begin{align}
    \min\;&
    J(\mbf{U}, \mbf{\nu}, \mbf{W}, \gamma, \mbf{V}_1, \mbf{V}_2; \beta)
    =
    c
    -2 \trace{\left(
        \mbf{U} \mbf{G}^\trans
    \right)}
    +
    \nu
    +
    \frac{\beta}{q} \gamma
    \\
    \mathrm{s.t.}\;&
    \trace{(\mbf{W})} < \nu,
    \\
    &\mbf{W} > 0,
    \\
    &\begin{bmatrix}
        \mbf{W} & \mbf{U} \mbf{L} \\
        \mbf{L}^\trans \mbf{U}^\trans & \mbf{1}
    \end{bmatrix}
    >
    0,
    \\
    &\trace{(\mbf{V}_1)} + \trace{(\mbf{V}_2)}
    \leq
    2\gamma,
    \\
    &\begin{bmatrix}
        \mbf{V}_1 & \mbf{U} \\
        \mbf{U}^\trans & \mbf{V}_2
    \end{bmatrix}
    \geq 0,
\end{align}
where $\mbf{H} = \mbf{L} \mbf{L}^\trans$,
is equivalent to minimizing~\eqref{eq:nuclear-reg-cost}.

\section{Asymptotic stability constraint}

To ensure that all eigenvalues associated with the matrix $\mbf{A}$, where
$\mbf{U} = \begin{bmatrix} \mbf{A} & \mbf {B}\end{bmatrix}$,
have magnitude strictly less than one, thus ensuring asymptotic stability,
a modified Lyapunov constraint~\cite[\S1.4.4]{ghaoui_advances_2000}
\begin{align}
    \mbf{P} &> 0,
    \\
    \mbf{A}^\trans \mbf{P} \mbf{A} - \bar{\rho}^2\mbf{P} &< 0,%
    \label{eq:modified-lyap}
\end{align}
can be added to ensure that the magnitude of the largest eigenvalue of $\mbf{A}$
is no larger than $0 < \bar{\rho} < 1$.
Applying the Schur complement to~\eqref{eq:modified-lyap} yields
\begin{align}
    \mbf{A}^\trans
    \mbf{P}
    \mbf{A}
    -
    \bar{\rho}^2
    \mbf{P}
    <
    0
    &\iff
    \left(%
        \mbf{A}^\trans
        \mbf{P}
    \right)
    \mbf{P}^{-1}
    {\left(%
        \mbf{A}^\trans
        \mbf{P}
    \right)}^\trans
    -
    \bar{\rho}^2
    \mbf{P}
    <
    0
    \\
    &\iff
    \left(%
        \mbf{A}^\trans
        \mbf{P}
    \right)
    {\left(\bar{\rho}\mbf{P}\right)}^{-1}
    {\left(%
        \mbf{A}^\trans
        \mbf{P}
    \right)}^\trans
    -
    \bar{\rho}
    \mbf{P}
    <
    0
    \\
    &\iff
    \begin{bmatrix}
        - \bar{\rho}\,\mbf{P} & \mbf{A}^\trans\mbf{P} \\
        \mbf{P}^\trans\mbf{A} & - \bar{\rho}\,\mbf{P}
    \end{bmatrix}
    <
    0,\ -\bar{\rho}\,\mbf{P} < 0
    \\
    &\iff
    \begin{bmatrix}
        \bar{\rho}\,\mbf{P} & \mbf{A}^\trans\mbf{P} \\
        \mbf{P}^\trans\mbf{A} & \bar{\rho}\,\mbf{P}
    \end{bmatrix}
    >
    0,\ \mbf{P} > 0.
\end{align}
The full optimization problem with asymptotic stability constraint is therefore
\begin{align}
    \min\;&
    J(\mbf{U}, \mbf{\nu}, \mbf{W}, \mbf{P}; \bar{\rho})
    =
    c
    -2 \trace{\left(
        \mbf{U} \mbf{G}^\trans
    \right)}
    +
    \nu
    \\
    \mathrm{s.t.}\;&
    \trace{(\mbf{W})} < \nu,
    \\
    &\mbf{W} > 0,
    \\
    &\begin{bmatrix}
        \mbf{W} & \mbf{U} \mbf{L} \\
        \mbf{L}^\trans \mbf{U}^\trans & \mbf{1}
    \end{bmatrix}
    >
    0,
    \\
    &\mbf{P} > 0,
    \\
    &\begin{bmatrix}
        \bar{\rho}\,\mbf{P} & \mbf{A}^\trans\mbf{P} \\
        \mbf{P}^\trans\mbf{A} & \bar{\rho}\,\mbf{P}
    \end{bmatrix}
    > 0,
\end{align}
where $\mbf{H} = \mbf{L} \mbf{L}^\trans$ and
$\mbf{U} = \begin{bmatrix} \mbf{A} & \mbf {B}\end{bmatrix}$.

Since both $\mbf{A}$ and $\mbf{P}$ are unknown, this optimization problem
is bilinear, and can be solved iteratively by holding either $\mbf{A}$ or
$\mbf{P}$ fixed while solving for the other. Iteration must be performed until
the cost function stops changing significantly.

\section{System norm regularization}

A system norm like the \Hinf{}~norm, the \Htwo{}~norm, or a mixed \Htwo{}~norm
can be used to regularize the Koopman regression problem when it is posed as in
Section~\ref{sec:lmi}.
The use of the \Hinf{}~norm as a regularizer when finding the Koopman matrix via
regression is explored next.
Minimizing the \Hinf{}~norm guarantees that the resulting LTI system will be
asymptotically stable, and allows the regularization problem to be tuned in the
frequency domain with weighting functions.

The Koopman representation of a nonlinear ODE can be
thought of as a discrete-time LTI system
$\mbs{\mc{G}}: \ell_{2e} \to \ell_{2e}$, where $\ell_{2e}$ is the extended
inner product sequence space~\cite{zhou_robust_1995},
$\mbf{U} = \begin{bmatrix}\mbf{A} & \mbf{B}\end{bmatrix}$,
$\mbf{C}=\mbf{1}$, and $\mbf{D}=\mbf{0}$. That is,
\begin{equation}
    \mbs{\mc{G}}
    \stackrel{\min}{\sim}
    \bma{c|c}
        \mbf{A} & \mbf{B} \\
        \hline 
        \mbf{C} & \mbf{D}
    \ema,
\end{equation}
where $\stackrel{\min}{\sim}$ denotes a minimal state space
realization~\cite[\S16.9.16]{bernstein_2018_scala}.
The $\mc{H}_\infty$ norm of $\mbs{\mc{G}}$ is the
worst-case gain from $\|\mbf{u}\|_2$ to $\|\mbs{\mc{G}}\mbf{u}\|_2$.
That is~\cite[\S3.2.2]{caverly_2019_lmi},
\begin{equation}
    \|\mbs{\mc{G}}\|_\infty
    =
    \sup_{\mbf{u} \in \ell_2, \mbf{u} \neq \mbf{0}}
    \frac{\|\mbs{\mc{G}}\mbf{u}\|_2}{\|\mbf{u}\|_2}.
\end{equation}

With \Hinf{} norm regularization,
the cost function associated with the regression problem is
\begin{equation}
    J(\mbf{U}; \beta)
    =
    \|\mbs{\Theta}_+ - \mbf{U} \mbs{\Psi}\|_\frob^2
    +
    \beta \|\mbs{\mc{G}}\|_\infty,
    \label{eq:hinf-reg-cost}
\end{equation}
where $\beta$ is the regularization coefficient.

The \Hinf{} norm has an LMI formulation. The inequality
$\|\mbs{\mc{G}}\|_\infty < \gamma$
holds if and only if~\cite[\S3.2.2]{caverly_2019_lmi}
\begin{align}
    \mbf{P} > 0,
    \\
    \begin{bmatrix}
        \mbf{P} & \mbf{A} \mbf{P} & \mbf{B} & \mbf{0} \\
        \mbf{P} \mbf{A}^\trans & \mbf{P} & \mbf{0} & \mbf{P} \mbf{C}^\trans \\
        \mbf{B}^\trans & \mbf{0} & \gamma \mbf{1} & \mbf{D}^\trans \\
        \mbf{0} & \mbf{C} \mbf{P} & \mbf{D} & \gamma \mbf{1}
    \end{bmatrix}
    >
    0.
\end{align}

The full optimization problem with \Hinf{} regularization is
\begin{align}
    \min\;&
    J(\mbf{U}, \mbf{\nu}, \mbf{W}, \gamma, \mbf{P}; \beta)
    =
    c
    -2 \trace{\left(
        \mbf{U} \mbf{G}^\trans
    \right)}
    +
    \nu
    +
    \frac{\beta}{q}\gamma
    \\
    \mathrm{s.t.}\;&
    \trace{(\mbf{W})} < \nu,
    \\
    &\mbf{W} > 0,
    \\
    &\begin{bmatrix}
        \mbf{W} & \mbf{U} \mbf{L} \\
        \mbf{L}^\trans \mbf{U}^\trans & \mbf{1}
    \end{bmatrix}
    >
    0,
    \\
    &\mbf{P} > 0,
    \\
    &\begin{bmatrix}
        \mbf{P} & \mbf{A} \mbf{P} & \mbf{B} & \mbf{0} \\
        \mbf{P} \mbf{A}^\trans & \mbf{P} & \mbf{0} & \mbf{P} \mbf{C}^\trans \\
        \mbf{B}^\trans & \mbf{0} & \gamma \mbf{1} & \mbf{D}^\trans \\
        \mbf{0} & \mbf{C} \mbf{P} & \mbf{D} & \gamma \mbf{1}
    \end{bmatrix}
    >
    0,
\end{align}
where $\mbf{H} = \mbf{L} \mbf{L}^\trans$ and
$\mbf{U} = \begin{bmatrix} \mbf{A} & \mbf {B}\end{bmatrix}$.

This is a bilinear optimization problem as both $\mbf{P}$ and $\mbf{A}$ are
unknown. It must be solved iteratively by holding either $\mbf{P}$ or $\mbf{A}$
fixed while solving for the other. Iteration must be performed until the cost
function stops changing significantly.

\section{Reproducible research}

The methods presented in this document are implemented in \texttt{pykoop}, the
authors' open source Koopman operator identification
library~\cite{dahdah_2021_pykoop}.

\section{Conclusion}

Regression is one way to approximate a Koopman matrix from data.
The presented LMI-based methods to regularize and constrain the Koopman matrix
regression problem are part of a modular approach that can be readily adjusted
for the problem at hand.
The proposed method of regularizing the Koopman matrix regression
problem with the \Hinf{}~norm provides a systems perspective to the problem and
allows the regularization to be tuned in the frequency domain using weighting
functions. Other system norms, like the \Htwo{}~norm, the generalized
\Htwo{}~norm, the peak-to-peak norm~\cite[\S3]{caverly_2019_lmi} can also be
used as regularizers. The unique properties of these system norms may prove
useful in the identification of approximate Koopman operators from data.
Further exploration of these properties, along with investigation into improved
methods to solve the bilinear matrix inequalities arising in these problems is
the subject of future research.

\printbibliography{}

\end{document}